\documentclass{aa}
\newcommand{\be}{\begin{equation}}
\newcommand{\ee}{\end{equation}}
\usepackage{graphicx}

\begin{document}

\title{Global spiral modes in multi-component disks}
\author{Orlova N.\inst{1},
    Korchagin V.\inst{1} and Theis Ch.\inst{2}}
  \institute{Institute of Physics, Stachki 194, Rostov-on-Don, Russia,
  email: nata@ip.rsu.ru \and
  Institut f\"ur Theoretische Physik und Astrophysik d.\ Univ.\ Kiel,
  D--24098 Kiel, Germany,  email: theis@astrophysik.uni-kiel.de}
\offprints{N.\ Orlova}
\date{received \ 22 June 2001; accepted \ 19 October 2001}
\abstract{
We performed two-dimensional non-linear
hydrodynamical simulations of two-component gravitating
disks aimed at studying stability properties of these systems.
In agreement with previous analytical and numerical simulations,
we find that the cold gas component
strongly affects the growth rates of the unstable global
spiral modes in the disk.
Already a five percent admixture of cold gas increases
approximately two-fold the growth rate of the most
unstable global mode while a twenty percent gas admixture
enhances the modal growth rate by a factor of four.
The local stability properties of a two-component disk
coupled by self-gravity are governed by a stability
criterion similar to Toomre's Q-parameter derived for one-component 
systems. Using numerical simulations, we analyse
the applicability of a two-component local stability criterion 
for the analysis of the stability properties
of global modes.
The comparison of non-linear simulations with the 
analytical stability criterion shows that
the two-component disks can be globally unstable while
being stable to the local perturbations. The minimum value of the
local stability criterion provides, however,
a rough estimate of the global stability properties of
two-component systems. 
We find that two-component systems with a content of cold gas 
of ten percent or less are globally stable
if the minimum value of the stability parameter exceeds $\sim$ 2.5.
}

\maketitle


\section{Introduction}

Global spiral arms observed in disk galaxies are the result of
density waves propagating in the galactic disks (e.g.\ Binney \& Tremaine 
\cite{binney87}).
Starting from pioneering works by B. Lindblad (\cite{lindblad41},
\cite{lindblad42}),
a number of authors extensively investigated this idea,
and a variety of techniques have been developed to study
the stability properties of gravitating disks. Most of these studies
concentrated on one-component gaseous or collisionless
stellar disks. 
Galactic disks, however, are multi-component systems
whose properties depend on all the constituents.
Lin \& Shu (1966) first addressed the question showing
the relative importance of gas in providing the spiral
gravitational field.
Jog \& Solomon (\cite{jog84}) (hereafter JS) 
considered the influence of a cold component on the
stability properties of gravitating disks.
They derived a local dispersion
relation for a two-fluid instability of a star-gas disk and
discussed a relative contribution of two components
in the stability properties of the disk. In further investigations
the results of JS were generalized by 
Bertin \& Romeo (1988) who studied the stability 
properties of the discrete global spiral modes by applying
asymptotic WKB-methods. They also introduced 
the concept of the effective $Q$-parameter of the two-component
disks.
Elmegreen (1995) re-considered the original
equations derived by JS and reduced them to a single effective
$Q$-parameter. Romeo (1992, 1994) included the effects of the
finite thickness of stellar and gaseous disks.
Recently, Rafikov (\cite{rafikov01}) gave the analysis
of the axisymmetric gravitational stability of the thin
rotating disk consisting of several components.

The non-asymptotic global modal stability properties of 
multi-component disks
were studied to a lesser extent. 
Noh et al.\ (\cite{noh91}) studied the
global stability properties of a
thin two-fluid protoplanetary disk composed of
gas and dust coupled by gravity and friction
and found that the stability properties of these disks are strongly affected
by a small amount of dust.
Kikuchi et al.\ (\cite{kikuchi97})
studied global stability properties of a three-phase disk
with phases coupled by self-gravity and interchange
processes between the phases. Korchagin \& Theis (\cite{korchagin99}) 
numerically studied the behavior of global modes in multi-component
disks at the non-linear stage of global instability.

In this paper, we return to the problem of the global
stability properties of multi-component disks.
We consider the simplest possible system of that kind -- a two-fluid
thin disk coupled by the self-gravity of the components.
We compare the results of our non-linear two-dimensional numerical
simulations with the local stability criteria derived by JS
and study the applicability of the Jog-Solomon criterion for the
prediction of the global stability properties of multi-phase
gravitating disks. We also study the non-linear behaviour of the
unstable global modes in multi-phase disks determining
the dependence of the saturation levels of the unstable global modes
on the mass fraction of a cold component.

Sect.\ \ref{sectbasic} describes the basic equations and
the numerical code used in simulations. 
In Sect.\ \ref{sectequilibriummodel} we present
the equilibrium model used in our simulations. In Sect.\ 
\ref{sectstabilitycrit}  
we briefly describe the two-fluid local stability criterion. 
Sect.\ \ref{sectresults} presents the results of our simulations and gives a 
comparison of the results of the numerical simulations with the two-fluid
analytical stability criterion.


\section{Basic Equations}
\label{sectbasic}

To model the behaviour of the global perturbations in
a gravitationally coupled multi-component disk
we use the hydrodynamic approximation, modelling both a stellar and a
gaseous component as polytropic fluids:
\be
P_{s,g} = K_{s,g} \sigma_{s,g}^{\gamma_{s,g}}
\label{eq1}
\ee
Here $P_{s,g}$ are the vertically integrated pressures of each
component, $K_{s,g}$
are the polytropic constants and $\gamma_{s,g}$ are the polytropic indices for
stellar and gaseous disks, respectively.
In this approach, stellar and gaseous `fluids' differ by the
values of their polytropic indices $\gamma_{s,g}$. We 
choose the value $\gamma_s = 2$ for the stellar disk which 
reproduces the empirical
``square root'' proportionality between the velocity dispersion
and the surface density found in galactic disks (Bottema \cite{bottema92}).
Additional arguments justifying the description of the
collisionless stellar disks by a fluid dynamical approach are given by 
Kikuchi et al.\ (\cite{kikuchi97}). The polytropic index for the gaseous disk 
is assumed to be $\gamma_g = 1.66$.

The behaviour of a two-fluid gravitationally interacting system
is described by the continuity and momentum equations for
each component: 
\begin{equation}
\label{eq2}
 {\partial \sigma_{s,g} \over \partial t}
 + {1 \over r} {\partial \over \partial r} \Big( r\sigma_{s,g} u_{s,g} \Big)
 + {1 \over r} {\partial \over \partial \phi} \Big( \sigma_{s,g} v_{s,g} 
     \Big) = 0
\end{equation}
\begin{eqnarray}
\label{eq3} 
{\partial u_{s,g} \over \partial t}
 + u_{s,g} {\partial u_{s,g} \over \partial r}
 + {v_{s,g} \over r} {\partial u_{s,g} \over \partial \phi}
 - {v^2_{s,g} \over r} = \nonumber \\
  - {1 \over \sigma_{s,g}} {\partial P_{s,g} \over \partial r} 
 - {\partial \over \partial r} \Big( \Psi + \Psi _H \Big)
\ ,
\end{eqnarray}
\begin{eqnarray}
\label{eq4}
{\partial v_{s,g} \over \partial t}
 + u_{s,g} {\partial v_{s,g} \over \partial r}
 + {v_{s,g} \over r} {\partial v_{s,g} \over \partial \phi}
 + {v_{s,g} u_{s,g} \over r} = \nonumber \\
 - {1 \over \sigma_{s,g} r} {\partial P_{s,g} \over \partial \phi}
 - {1 \over r} {\partial \over \partial \phi} \Big( \Psi + \Psi_H \Big)
\ .
\end{eqnarray}
\noindent
Additionally, all components are coupled by their gravity, which can be
expressed by the Poisson equation
\begin{equation}
\Psi(r,\phi) = -
\int_{R_{\rm in}}^{R_{\rm out}} \int_{0}^{2 \pi}
{(\sigma_s(r^{\prime},\phi^{\prime}) + \sigma_g(r^{\prime},\phi^{\prime}))
 r^{\prime} dr^{\prime} d \phi^{\prime}
 \over {\sqrt{r^{2}+r^{\prime 2} -
 2rr^{\prime} \cos (\phi - \phi^{\prime})}}}
\ .
\label{eq5}
\end{equation}
Here $u_{s,g}$ and $v_{s,g}$ are the radial 
and azimuthal velocities of stars and gas within the disk, and
$\sigma_{s,g}$ is the surface density of each component. 
The potential $\Psi$ derived from the self-gravity of a 
two-fluid disk, 
the explicit contribution $\Psi_H$ from
the rigid halo, and the pressure gradient determine the behaviour of
the disk.
All of the dependent variables are functions of the radial coordinate
$r$, the azimuthal angle $\phi$, and the time $t$.
The variables and the parameters of the model are
expressed in units in which
the gravitational constant  $G$ is equal to unity,
the unit of mass is equal to $10^{10}M_{\odot}$
and the unit of length is $2$ kpc. This choice 
determines the velocity unit $V_{\rm unit}=149.1$ \mbox{km\,s$^{-1}$} and the 
time unit $T_{\rm unit}=1.32 \times 10^7$ yr.

To solve the two-component hydrodynamical equations (\ref{eq2}) -- (\ref{eq4})
we use the multi-component fluid-dynamical code developed
by Korchagin \& Theis (\cite{korchagin99}). The code is a multi-phase
realization of the Eulerian ZEUS-2D code (Stone \& Norman \cite{stone92})
with the Van-Leer advection scheme.
The two-fluid hydrodynamical equations are solved using
equally spaced azimuthal and logarithmically spaced radial
zones. In the numerical simulations discussed here we
employed a grid of 256 $\times$ 256 cells. The Poisson
equation is solved by the 2D Fourier convolution
theorem in polar coordinates.
We look for the global modes starting from initial random
perturbations at the level $10^{-6}$.


\section{Equilibrium model}
\label{sectequilibriummodel}

The equilibrium disk taken in our simulations is based  
on the equilibrium properties of real galaxies. 
Specifically, we applied the model of the spiral galaxy NGC 1566 
by Korchagin et al.\ (\cite{korchagin00}). The 
disk of NGC 1566 is observationally well studied.
Its rotation curve
taken from Bottema (\cite{bottema92}) can be approximated by
\begin{equation}
v_0(r) = {V_1 r \over (r^2 + R_1^2)^{3/4}}
       + {V_2 r \over (r^2 + R_2^2)^{3/4}} 
\ .
\label{eq6}
\end{equation}
The radial dependence of the vertical velocity dispersion of the stars
in the disk of NGC 1566 has an exponential distribution 
\be
c_z = c_z(r) = c_{z0} \exp(-r/h_{c_z}) {\mbox .}
\label{eq7}
\ee
This uniquely determines the radial dependence of the
axisymmetric surface density distribution assuming
that the disk is self-gravitating in the vertical direction:
\be
\sigma(r) = c_z^2(r) /\pi G z_0 {\mbox .}
\label{eq8}
\ee
Here $G$ is the gravitational constant, and $z_0$ is the effective
thickness of the disk.
According to equations (\ref{eq7}) and (\ref{eq8}), 
the radial distribution of the stellar surface density can be expressed by
\be
\sigma_{s}(r) = \sigma_{0_{s}} \exp(-r/h_{\sigma})
              [1-(r/R_{\rm out})^2]^5{\mbox ,}
\label{eq9}
\ee
where $h_{\sigma}$ is radial scale length for the stellar surface density 
distribution. For numerical purposes
we use in equation (\ref{eq9}) a multiplier $[1-(r/R_{\rm out})^2]^5$ 
which results in a vanishing surface density at the outer boundary of the disk.
We further assume that the surface density profile
of the gaseous disk and its thickness are the same as those in 
the stellar disk.
However, the total mass of the gaseous disk is smaller compared to the
mass of the stellar disk and varies in our models.

The radial velocity dispersion of the
stellar and gaseous components can be derived from the equation of state by
\be
c_{r_{s,g}} = (\gamma_{s,g} K_{s,g} \sigma_{s,g}(r)^{\gamma_{s,g}-1})^{1/2}
\mbox{\,\, .}
\label{eq10}
\ee
The free parameters $K_{s,g}$ can be fixed by the equations (\ref{eq7}) and
(\ref{eq8}) and by the 
assumption of a constant ratio of the vertical and radial velocity 
dispersions. For the latter, to get the disk with
weak global modes we have
chosen $c_{sz}/c_{sr} = 0.9$. 
Our choice can be justified as follows. 
According to Bottema (1992), the measured velocity dispersion in    
the disk of NGC 1566 gives the real dispersion perpendicular
to the galactic disk. In our one-component model, we fix the
$z$-component of the velocity dispersion and hence the surface
density and the total mass of the disk, and allow for a variation of the
radial velocity dispersion. Increasing the ratio $c_{sz}/c_{sr}$
we decrease the radial velocity dispersion, hence
destabilizing the disk.                
Simulations show that for the measured $z$-dispersion and rotation
curve in the disk of NGC 1566, the disk is stable for the usually
assumed ratio $c_{sz}/c_{sr} = 0.6$. Therefore, we choose in our
studies the ratio $c_{sz}/c_{sr} = 0.9$ which leads to a weakly
unstable disk. This ratio is larger than the measured
ratio of the velocity dispersions $c_{sz}/c_{sr} = 0.5 - 0.6$
in solar neighborhood, and close to the upper limit      
measured in other galaxies.
E.g., the determination of this ratio                            
in the disk of galaxy NGC 488 undertaken by  Gerssen et al. (1997)
gives the value $0.7 \pm 0.19$. With our choice, however,
we do not aim to model any particular galaxy.

The equilibrium rotation of the disk given by the equation (\ref{eq6})
is balanced by the self-gravity, the pressure gradient of the disk and 
the external potential of a rigid halo:
\be
{v_0^2 \over r}
   = {1 \over \sigma} {d P_s \over d r}
   + {d \over dr} \Big( \Psi + \Psi_H \Big )
\ .
\label{eq11}
\ee

   \begin{table}
      \caption{Parameters of the equilibrium model
                 used in the numerical simulations}
         \label{table1}
         \begin{tabular}{ll}
            \hline
            Inner disk radius $(R_{\rm in})$, kpc           & 0.2   \\
            Outer disk radius $(R_{\rm out})$, kpc          & 10.0  \\
            Disk thickness $(z_0)$, kpc                     & 0.7   \\
            Scale length of velocity dispersion $(h_{c_z})$, kpc & 2.6  \\
            Scale length of surface density $(h_{\sigma})$, kpc  & 1.3  \\
            Central velocity dispersion $(c_{z_0})$, km\,s$^{-1}$ & 168 \\
            Total mass of the disk $(M_d)$, $M_{\odot}$ & $1.78 \times 10^{10}$\\            Rotation curve parameters:       &                            \\
            $V_1$, $V_2$, km\,s$^{-1}$       & 399, 82  \\
            $R_1$, $R_2$, kpc                & 4.4, 0.6 \\
            \hline
         \end{tabular}
\end{table}

By this, we can fix the halo properties.
Knowing the initial surface density of the disk
we get the disk's self-gravity by solving
the Poisson equation numerically. Equation (\ref{eq11})
can then be used to calculate the gradient of the
halo potential and the halo mass distribution.
The general parameters of our equilibrium model
are listed in Table \ref{table1}.

\section{Stability Criteria}
\label{sectstabilitycrit}

The local stability of a thin gravitating disk can be
expressed in terms of Toomre's $Q$ parameter which for the
gaseous disk has the form (Safronov \cite{safronov60}, 
Toomre \cite{toomre64}):
\begin{equation}
Q \equiv {c(r) \kappa(r) \over \pi G \sigma(r)}
\ .
\label{eq12}
\end{equation}
Here, $c(r)$ is the sound speed of the gas, $\kappa(r)$ is the
epicyclic frequency and $\sigma(r)$ is the surface density of the disk.
A gravitating disk is locally unstable when $Q < 1$.

The stability criterion (\ref{eq12}) was derived for
tightly wound perturbations. However, it can be successfully
used as an indicator for the stability of global modes
in gravitating disks. Numerical simulations show that a locally stable
stellar or
gaseous disk can be globally unstable if the minimum value of
$Q$ does not exceed $ \sim 2$.
For example, a one-component disk with the rotation and surface
density distribution given by equations (\ref{eq6}) and (\ref{eq9})
can be globally unstable, if the minimum value of $Q$
does not exceed $\sim$ 1.6 (Korchagin et al. \cite{korchagin00}).

Disks of real galaxies consist of stellar and gaseous components.
The stellar component comprises the largest fraction
of the total mass of the disk. However, colder gaseous components
can play a significant role for the stability properties
of multi-component disks.
JS derived a local stability criterion for a two-component fluid
disk which can be written in the form $Q_{\rm JS} < 1$
where

\be
 Q_{\rm JS} = 
\Big( {2 \over Q_s}{q \over 1 + q^2} + {2 \over Q_g}{c_g \over c_s}
{q \over 1 + q^2c_g^2/c_s^2} \Big)^{-1}  
\ .
\label{eq13}
\ee

Here $Q_s$ and $Q_g$ are Toomre's $Q$-parameters determined
for the `stellar' and gaseous fluids, correspondingly.
$q$ is the dimensionless radial wavenumber of perturbations
$q = kc_s/\kappa$.
The relative contribution $\Gamma$ of gas to stars is given by the ratio
of the second to the first term of the stability parameter (\ref{eq13}):
\be
 \Gamma = {\sigma_g \over \sigma_s} {1 + q^2 \over 1 + q^2c_g^2/c_s^2}
\ .
\label{eq14}
\ee

It can be seen from equation (\ref{eq13}) that the relative contribution
of the components is always larger than the ratio of the surface
densities of the components, provided the gas is dynamically colder than
the stars.
In the limiting case of large wavenumbers $q \rightarrow \infty$  
the contribution of the components
is given by the ratio $\sigma_g c_s^2 / \sigma_s c_g^2$ - the
result, obtained by Marochnik \& Suchkov (\cite{marochnik74}).

We will use the criterion given by equation (\ref{eq13})
for the analysis of instability of the global modes in
two-component gravitating disks.

\begin{figure}
   \centering
   \includegraphics[angle=-90,width=9.5cm]{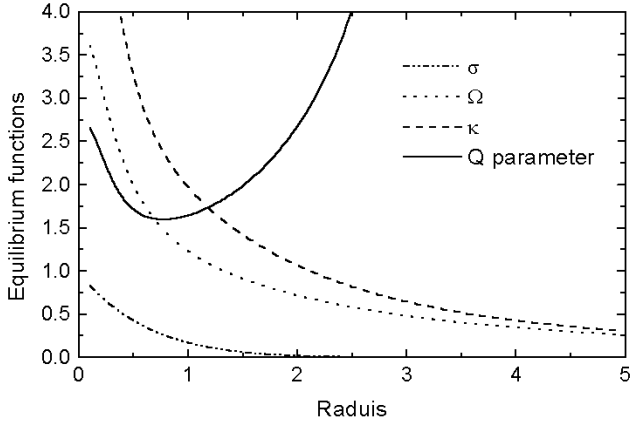}
   \caption{
           Equilibrium curves for the one-component `stellar' disk.
           Angular velocity $\omega$ and epicyclic
           frequency $\kappa$ are given in units of
           $0.75 \times 10^{-7}{\rm yr}^{-1}$.
           The surface density $\sigma$ is given in units
           $2.5 \times 10^9 M_{\odot} {\rm kpc}^{-2}$,
           radius is in units of length $L=2$ kpc.}
      \label{fig1}
\end{figure}


\section{Results}
\label{sectresults}

To study the influence of a cold component on the overall dynamics
of a two-fluid gravitating disk, we start our analysis with a one-component
`stellar' disk which we choose to be close to marginal stability.
Fig.\ \ref{fig1} shows the equilibrium properties of this disk. The local
stability parameter $Q$ is well above unity with a
minimum value $Q_{\rm min} \sim 1.6$. 


The radial behavior of the $Q$-parameter is one of the
key assumptions in the modal analyses of spiral density waves. 
In previous studies it was usually assumed that the profile
of the stability parameter $Q$ has a barrier at the
central regions of the disk and is equal to unity in the
outer regions of the disk (e.g.\ Bertin et al.\ 1989). 
Such a behavior of an `effective'
$Q$-parameter is supposed to model
the absence of cold gas as well as the transition in geometry
from the disk to the nuclear bulge in the central regions
of a galaxy, and to describe the increasing role of a cold gaseous component
in the outer regions of a galactic disk.
In this paper we do not introduce an effective $Q$-parameter, but
we try to base our studies on observational properties
of disk galaxies. We choose
the basic state in our fiducial model
based on the properties of the
nearby late type spiral galaxy NGC 1566.
The `stellar' $Q$-parameter 
shown in Fig.\ \ref{fig1} has an outer barrier which is typical
for disks with exponentially decreasing radial surface
density distributions and exponentially decreasing 
velocity dispersions.

\begin{figure}
   \centering
   \includegraphics[angle=-90,width=9.5cm]{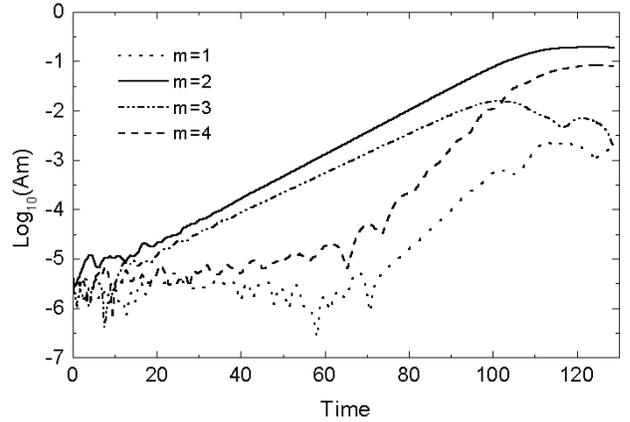}
   \caption{
          Temporal evolution of the $m=1-4$
          global Fourier amplitudes computed for the
          one-component `stellar' disk of Fig.\ \ref{fig1}.
          Time is in units of $1.32 \times 10^7$ yr.}
       \label{fig2}
\end{figure}

Linear global modal analysis of this model reveals a few slowly 
growing modes with
the fastest growing spiral $m=2$.
Nonlinear 2D simulations
confirm this result giving an exponential growth rate of the $m = 2$ mode 
of approximately 0.11 in dimensionless units. Linear theory predicts a
value of 0.12 which is in good agreement with non-linear simulations.
Fig.\ \ref{fig2} illustrates the
evolution of the $m = 1 - 4$ global modes in
the one-component `stellar' disk given in terms of the   
global Fourier amplitudes
\be
A_m \equiv {1 \over {M_{d}}}
\left\vert \int_{0}^{2 \pi} \int_{R_{\rm in}}^{R_{\rm out}}
\sigma(r,\phi) r dr \, e^{-im\phi} d\phi \right\vert \mbox{\,\, .}
\label{eq15a}
\ee
The most unstable $m=2$ mode grows linearly until
$t \approx 110$, and then saturates at a level $A_2 \approx 0.2$ 
(Fig.\ \ref{fig2}). The linear regime
corresponds to an exponential growth of perturbations.

 In the next step we split
the one-component disk into a `stellar' and a `gaseous'
subsystem keeping the total mass of the disk constant.
We consider a set of models with a cold component 
consisting of 5\%, 10\% and 20\% of the total
mass of the disk.
In the first series of simulations we keep the `stellar' and
gaseous equations of state
the same for all the models. As a result,
the gas velocity dispersion, which depends on the gas fraction,
is not constant.
Thus, the velocity dispersion of gas is about 20\%
of the velocity dispersion of the stars in the model with 5\% gas admixture.
This ratio increases up to 36\% for the model with 20\% gas admixture. 

The strong influence
of the gas fraction is shown in Fig.\ 3 which displays the growth rate of
the most unstable $m=2$ mode in a one-component `stellar'
disk compared to the modal growth rate of the two-component models 
with 5\%, 10\%, and 20\%
gas admixtures. Already a five percent gas
fraction enhances the modal growth rate by a factor of 1.7 compared
to the one-component disk. A 20\% gas admixture results in an increase of
the growth rate by a factor of 3.7.
Table \ref{table2} gives the masses and the
dispersion ratio of the components in the computed models
together with the growth rates of the most unstable modes,
saturation levels and the minimum values of the
two-component stability criterion $Q_{\rm JS}$ calculated by 
equation (\ref{eq13}).
The growth rate of the dominant mode increases in the
models with higher gas content,
being the same for the gaseous and the stellar components. 
However, the saturation level of the most unstable mode is systematically
higher for the gaseous disk. 
In general, the saturation level   
slightly decreases in the models with higher gas content.
A possible explanation of this behavior is that
spiral shocks play a more important role in the saturation
of exponentially growing modes when more cold gas is present.
\begin{figure}
   \centering
   \includegraphics[angle=-90,width=9.5cm]{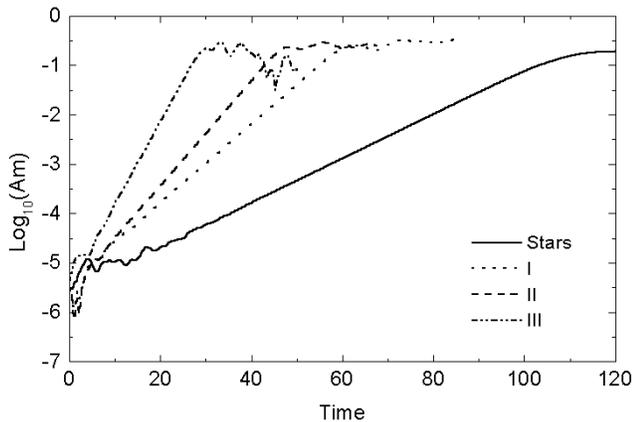}
   \caption{
          Time dependence of the global Fourier amplitudes
          for the most unstable $m=2$ spiral mode computed
          for models with different gas fractions 
          (I: 5\%, II: 10\%, III: 20\%).
          Time is in units of $1.32 \times 10^7$ yr.}
       \label{fig3}
\end{figure}
\begin{center}
\begin{table}
\caption[]{}
         \label{table2}
\begin{tabular}{c|c c c c c c}
\hline    &     &  &  & \multicolumn{2}{c}{Saturation level} &   \\
      Model &$M_g/M_s$ &$c_g/c_s$ &${\rm Im}_2$
            & gas &  stars & ${Q_{\rm JS}}_{\rm min}$ \\
\hline I   & 0.05  & 0.2 & 0.18 & 0.35 & 0.27 & 2.2  \\
       II  & 0.1 & 0.25 & 0.24 & 0.32 & 0.25 & 1.75 \\
       III & 0.2 & 0.36 & 0.4  & 0.29 & 0.22 & 1.36 \\
\hline
\end{tabular}
\begin{list}{}{}
  \item[Note.]     Dependence of the growth rates and the saturation
               levels on the content of a cold component for the
               governing $m=2$ mode.
               The total mass of the disk is fixed to
               $1.78 \times 10^{10} M_{\odot}$. The growth rates are given
               in units of $0.75 \times 10^{-7} {\rm yr}^{-1}$.
\end{list}
\end{table}
\end{center}
The ratio of the velocity dispersions of the subsystems
can also affect the stability of the disk. The influence of the
velocity dispersions on the growth rates of the global modes
was studied in a series of numerical experiments in which
the fraction of the cold component was fixed 
to 10\% of the total mass of the disk, while
the velocity dispersion ratio was allowed to change.
The results of these simulations are shown in Fig. 4.
The  growth rate of the most unstable global mode is only
slightly affected by the ratio of the velocity dispersions.
The saturation amplitude remains approximately at the same level.
Therefore, the mass fraction of the cold component is a more important
parameter for the growth rate of perturbations than the ratio of the
velocity dispersions. 
This is in agreement with results of Bertin \& Romeo (1988). They
demonstrated (see their Fig. 5) that the
criterion of marginal stability of a two-component disk also
depends slowly on the ratio of the velocity dispersions of the
components and more strongly on the ratio of their densities.
 
Disks with gaseous admixture show qualitatively the same 
evolution as the pure stellar disk. However, they evolve on a shorter 
time scale. The appearance of the global modes in all models 
is quite similar despite
the considerable difference in their growth rates.
Fig.\ \ref{fig5} shows the logarithmically 
spaced surface density contour plots
taken at the moments when the global amplitude of the $m=2$ mode in
all four models is $A_2 \approx 0.06$. All the models
develop an open two-armed spiral of comparable winding.
Note, however, that the spiral arms are more narrow, more tightly wound
and less extended when more gas is present.
Analysis of the global modes with help of the local dispersion relation
(Bertin \& Romeo 1988) also shows a shift towards tighter spirals with
a smaller corotation radius in models with higher gas content.

\begin{figure}
   \centering
   \includegraphics[angle=-90,width=9.5cm]{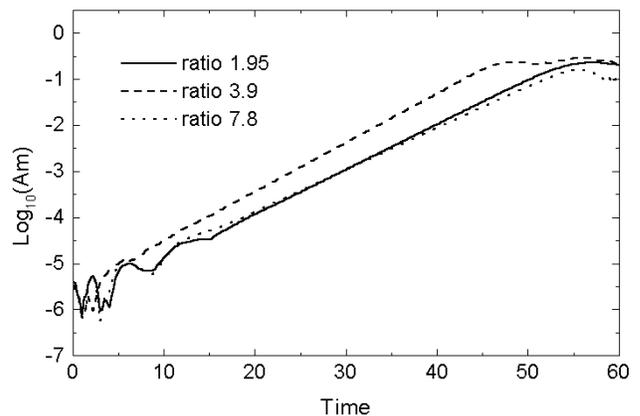}
   \caption{
           Time evolution of the global Fourier amplitude
           for the most unstable mode ($m=2$) computed for models
           with $10\%$ gas admixture and a ratio ${c_r}_s/{c_r}_g$ 
           of the velocity dispersions equal to 1.95, 3.9 and 7.8.
           Time is in units of $1.32 \times 10^7$ yr.}
         \label{fig4}
\end{figure}

\begin{figure*}
  \centering
  \includegraphics[angle=270,width=16cm]{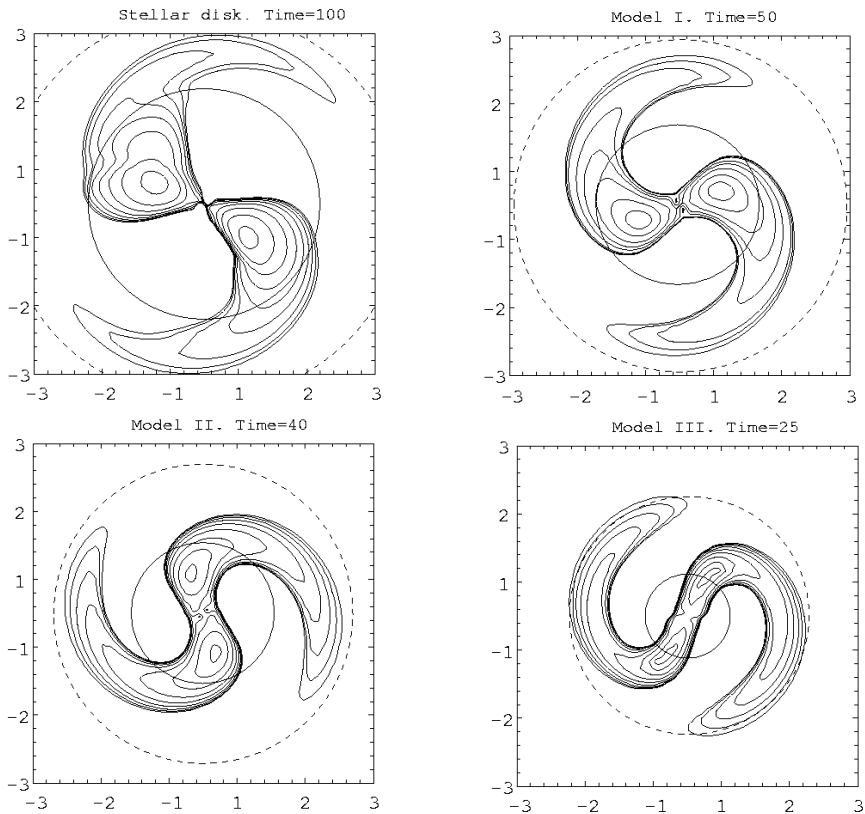}
   \caption {Surface density perturbations of the stellar component.
Displayed are the spiral perturbations in the one-component model
(upper left), and in the two-component models with different
gas fractions.
The snapshots
are taken at the moment when the $m=2$
global amplitude reaches $A_2 \approx 0.06$. 
The contour levels are logarithmically
spaced between the maximum value of the density perturbation
and one-hundredth of the maximum value. Circles show the positions
of the outer Lindblad resonance (dashed line) and corotation (solid line).}
       \label{fig5}
\end{figure*}

\begin{figure}
   \centering
   \includegraphics[angle=270,width=9.5cm]{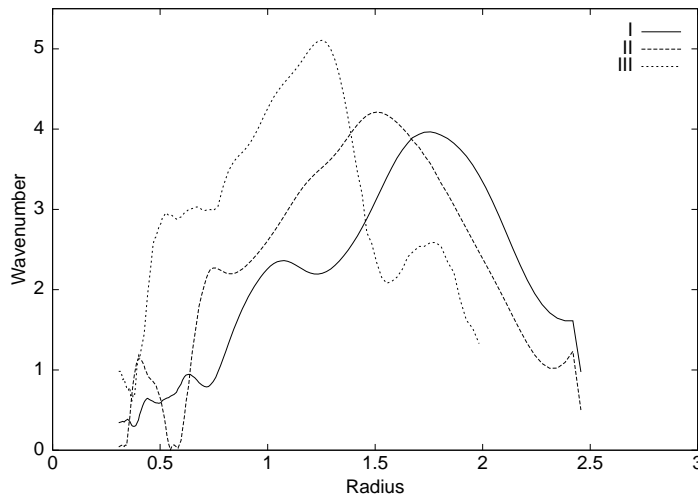}
   \caption{
          The radial dependence of the radial wavenumber of
          two-armed spiral for the models I - III.
          Unit of length $L=2$ kpc. }
       \label{fig6}
\end{figure}

To use the stability criterion (\ref{eq13}), one needs to know the wavenumber
of the perturbations. Fig.\ \ref{fig6} shows the radial dependence of  
the radial wavenumber for $m = 2$ perturbations computed 
for the models I - III. The wavenumbers were determined numerically
by calculating the pitch angles of the spiral arms shown in Fig.\ \ref{fig5}
at different radii, i.e.\ we used the relation
\begin{equation}
\cot i = \left\vert k R \over m \right\vert
\ .
\label{eq15b}
\end{equation}

To find a pitch angle of the spiral at a particular radius, 
we determine the line of the maximum amplitudes of the spiral arm and calculate
the angle between the tangents to the line of maxima and
the corresponding circle. 

\begin{figure}
   \centering
   \includegraphics[angle=270,width=9.5cm]{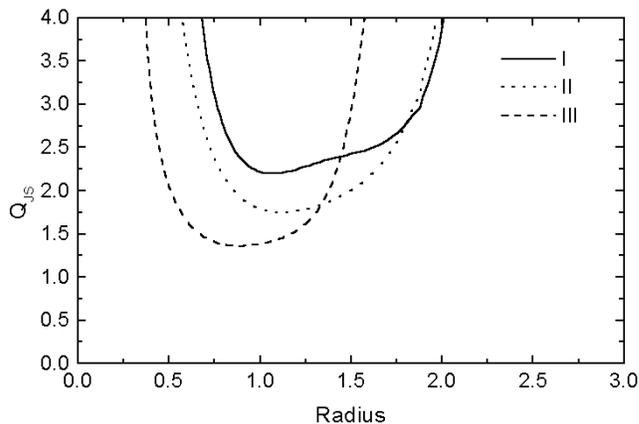}
   \caption{
          The radial dependence of the local two-fluid
          stability parameter $Q_{\rm JS}$
          for disk models with different gas fraction.
          Unit of length $L=2$ kpc.}
       \label{fig7}
\end{figure}

Fig.\ \ref{fig7} shows the radial profile of the dimensionless Jog-Solomon
stability parameter $Q_{\rm JS}$ computed with the wavenumbers
taken from Fig.\ \ref{fig6}. 
Fig.\ \ref{fig7} further illustrates
the destabilizing role of a cold gas component: the minimum
value $Q_{\rm JS}$ 
of the Jog-Solomon parameter decreases with increasing
content of cold gas. The minimum value
of the Jog-Solomon parameter $Q_{\rm JS}$  is about
1.36 for the two-component system with 20\% gas
admixture versus a minimum value of 2.2 found
for the two-component model with 5\% gas fraction.
In all three models, the Jog-Solomon stability criterion
exceeds unity, and all the models are 
locally stable. Such a behaviour is quite similar to the
behaviour of one-component disks which are known
can be globally unstable even if the minimum value of
Toomre's $Q$ parameter is above unity.


%
 
\begin{figure}
   \centering
   \includegraphics[angle=-90,width=9.5cm]{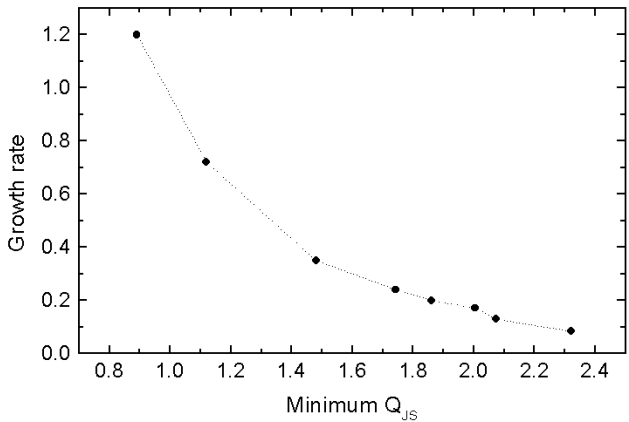}
   \caption{
          The growth rate as a function of the
          minimum value of local stability parameter $Q_{\rm JS}$.}
       \label{fig8}
\end{figure}

Decreasing the
total mass of the disk results in its stabilization which consequently leads
to an overall growth of the Jog-Solomon parameter $Q_{\rm JS}$.
Fig.\ \ref{fig8} 
illustrates this by displaying the growth rate of
the dominant mode $m=2$ as a function of the minimum value of
the Jog-Solomon parameter $Q_{\rm JS}$. The latter was obtained from numerical 
simulations of two-component disks with 10\% gas admixture.
These disks are stabilized when the minimum value of the Jog-Solomon
parameter exceeds $\sim 2.5$.

\section{Summary}

In this paper we studied the applicability of the two-fluid
local stability criterion derived by Jog \& Solomon (\cite{jog84}) 
for the onset and growth of non-axisymmetric perturbations in two-component 
gravitating disks with an open spiral structure. 
We performed three sets of numerical 
experiments: one set with fixed total mass of the disk and fixed 
polytropic equation of state, but variable gas fraction. Another set
of simulations was performed with a
fixed total mass of the disk and a fixed ratio of the stellar and 
gaseous velocity dispersions. A final set of calculations deals with 
a variable disk mass. 
Our numerical experiments have shown that the mass contribution
of the cold component strongly influences the stability properties
of the disk. Even a low gas fraction of $5\%$ destabilizes the disk
substantially, e.g.\ the growth rate of $m=2$ perturbations increases 
by a factor 1.7. All our disk models are unstable to $m=2$ spiral
modes. More unstable disks develop narrower and more
tightly wound spiral arms.
The saturation levels are almost constant for all simulations, with a 
slightly enhanced level for the gaseous component.
A comparison of the results of the direct 
two-dimensional numerical simulations with the JS stability
criterion allows us to conclude that similarly to Toomre's
stability parameter $Q$ derived for one-component disks,
the Jog-Solomon stability parameter $Q_{\rm JS}$ can be
used for the prediction of the global stability properties
of two-component gravitating disks which greatly simplifies
the analysis of disk stability. 


\begin{acknowledgements}

We thank the referee Dr. A.B. Romeo for his valuable comments on this work.
NO and VK gratefully acknowledge support by the
{\it Deutsche Forschungsgemeinschaft} under grant RUS 17/75/00(S).
\end{acknowledgements}

\end{document}